\documentclass[aps,pre,reprint]{revtex4-1}

\usepackage{amsmath}
\usepackage{amssymb}
\usepackage{marvosym}

\usepackage{graphicx}

\usepackage{color} 

\usepackage{hyperref}

\begin{document}


\title{Exact hybrid particle/population simulation of rule-based models of biochemical systems}

\author{Justin S.\ Hogg$^\textrm{\Yingyang}$}
\email[]{justinshogg@gmail.com}
\affiliation{Department of Computational and Systems Biology,
University of Pittsburgh School of Medicine,
Pittsburgh, PA 15260.}

\author{Leonard A.\ Harris$^\textrm{\Yingyang}$}
\email[]{lh64@cornell.edu}
\affiliation{Department of Computational and Systems Biology,
University of Pittsburgh School of Medicine,
Pittsburgh, PA 15260.}

\author{Lori J.\ Stover}
\affiliation{Department of Computational and Systems Biology,
University of Pittsburgh School of Medicine,
Pittsburgh, PA 15260.}

\author{Niketh S.\ Nair}
\affiliation{Department of Computational and Systems Biology,
University of Pittsburgh School of Medicine,
Pittsburgh, PA 15260.}

\author{James R.\ Faeder}
\email[]{faeder@pitt.edu}
\affiliation{Department of Computational and Systems Biology,
University of Pittsburgh School of Medicine,
Pittsburgh, PA 15260.}

\collaboration{{\footnotesize\Yingyang~These authors contributed equally to this work.}}

\begin{abstract} 
Detailed modeling and simulation of biochemical systems is complicated by the problem of combinatorial complexity, an explosion in the number of species and reactions due to myriad protein-protein interactions and post-translational modifications. Rule-based modeling overcomes this problem by representing molecules as structured objects and encoding their interactions as pattern-based rules. This greatly simplifies the process of model specification, avoiding the tedious and error prone task of manually enumerating all species and reactions that can potentially exist in a system. From a simulation perspective, rule-based models can be expanded algorithmically into fully-enumerated reaction networks and simulated using a variety of network-based simulation methods, such as ordinary differential equations or Gillespie's algorithm, provided that the network is not exceedingly large. Alternatively, rule-based models can be simulated directly using particle-based kinetic Monte Carlo methods. This ``network-free" approach produces exact stochastic trajectories with a computational cost that is independent of network size. However, memory and run time costs increase with the number of particles, limiting the size of system that can be feasibly simulated. Here, we present a hybrid particle/population simulation method that combines the best attributes of both the network-based and network-free approaches. The method takes as input a rule-based model and a user-specified subset of species to treat as population variables rather than as particles. The model is then transformed by a process of ``partial network expansion" into a dynamically equivalent form that can be simulated using a population-adapted network-free simulator. The transformation method has been implemented within the open-source rule-based modeling platform \mbox{BioNetGen}, and resulting hybrid models can be simulated using the particle-based simulator \mbox{NFsim}. Performance tests show that significant memory savings can be achieved using the new approach and a monetary cost analysis provides a practical measure of its utility. 
\end{abstract}

\pacs{}

\maketitle

\section*{Author Summary} 

Rule-based modeling is a modeling paradigm that addresses the problem of combinatorial complexity in biochemical systems. The key idea is to specify only those components of a biological macromolecule that are directly involved in a biochemical transformation. Until recently, this ``pattern-based" approach greatly simplified the process of model \emph{building\/} but did nothing to improve the performance of model \emph{simulation\/}. This changed with the introduction of ``network-free" simulation methods, which operate directly on the compressed rule set of a rule-based model rather than on a fully-enumerated set of reactions and species. However, these methods represent every molecule in a system as a particle, limiting their use to systems containing less than a few million molecules. Here, we describe an extension to the network-free approach that treats rare, complex species as particles and plentiful, simple species as population variables, while retaining the exact dynamics of the model system. By making more efficient use of computational resources for species that do not require the level of detail of a particle representation, this hybrid particle/population approach can simulate systems much larger than is possible using network-free methods and is an important step towards realizing the practical simulation of detailed, mechanistic models of whole cells.

\section*{Introduction}

\subsection*{Rule-based modeling}

Cell signaling encompasses the collection of cellular processes that sample the extracellular environment, process and transmit that information to the interior of the cell, and regulate cellular responses. In a typical scenario, molecules outside of the cell bind to cognate receptors on the cell membrane, resulting in conformational changes or clustering of receptors. A complex series of protein binding and biochemical events then occurs, ultimately leading to the activation or deactivation of proteins that regulate gene expression or other cellular processes \cite{Alberts02}. A typical signaling protein possesses multiple interaction sites with activities that can be modified  by direct chemical modification or by the effects of modification or interaction at other sites. This complexity at the protein level leads to a combinatorial explosion in the number of possible species and reactions at the level of signaling networks \cite{Hlavacek2003}. 

Combinatorial complexity poses a major barrier to the development of detailed, mechanistic models of biochemical systems. Traditional modeling approaches that require manual enumeration of all potential species and reactions in a network are infeasible or impractical \cite{Hlavacek2003, Hlavacek2006, Aldridge2006}. This has motivated the development of rule-based modeling languages, such as the \mbox{BioNetGen} language (BNGL) \cite{Blinov2004, Faeder2009}, Kappa \cite{Danos2004, Danos2007a}, and others \cite{Maus2011, Bittig2011, Lis2009, Angermann2012}, that provide a rich yet concise description of signaling proteins and their interactions \cite{Faeder2011}. The combinatorial explosion problem is avoided by representing interacting molecules as structured objects and using pattern-based rules to encode their interactions. In the graph-based formalisms of BNGL and Kappa, molecules are represented as graphs and biochemical interactions by graph-rewriting rules. Rules are \emph{local\/} in the sense that only the properties of the reactants that are transformed, or are required for the transformation to take place, affect their ability to react. As such, each rule defines a class of reactions that share a common set of transformations (e.g., the formation of a bond between molecules) and requirements for those transformations to take place (e.g., that one or more components have a particular covalent modification). The number of reactions encoded by a rule varies depending on the specifics of the model; a rule-based encoding is considered compact if it contains rules that encode large numbers of reactions. Overviews of rule-based modeling with BNGL can be found in Sec.~S3.1 of Text~S1 and Refs.~\cite{Faeder2009, Sekar2012}. A description of the graph-theoretic formalism underlying BNGL is provided in Sec.~S4.1 of Text~S1, building on a previous graph-theoretical treatment \cite{Blinov2006b}.

\subsection*{Network-based and network-free simulation of rule-based models}

An important characteristic of rule-based models is that they can encode both finite and \emph{infinite\/} reaction networks. If the network is finite and ``not too large" ($\lesssim$10\,000 reactions \cite{Sneddon2011}) it can be generated from the rule-based model algorithmically by a process known as ``network generation" \cite{Blinov2004, Faeder2005, Blinov2006b, Faeder2009, Sekar2012}. Network generation begins by applying the rules of a rule-based model to a set of initial ``seed" species, which define the initial state of the model system, to generate new species and reactions. The new species are then matched against the existing species to determine whether or not they are already present in the network \cite{Blinov2006a}. Any species that are not already present are added to the network and an additional round of rule application is performed. This iterative process continues until an iteration is encountered in which no new species are generated. The resulting system of reactions can then be simulated using a variety of network-based deterministic and stochastic simulation methods. For example, network-based simulation methods currently implemented within \mbox{BioNetGen} include SUNDIALS CVODE \cite{Hindmarsh2005} for ordinary differential equation (ODE)-based simulations, Gillespie's stochastic simulation algorithm (SSA; direct method with dynamic propensity sorting) \cite{Gillespie1976, Fricke1995}, and the accelerated-stochastic ``partitioned-leaping algorithm" \cite{Harris2006}.

The rule-based methodology also provides a way to simulate models with prohibitively large or infinite numbers of species and reactions. This ``network-free" approach involves representing molecular complexes as particles and applying rule transformations to those particles at runtime using a kinetic Monte Carlo update scheme \cite{Yang2008, Danos2007b}. At each simulation step, reactant patterns are matched to the molecular complexes within the system to calculate rule propensities. The rule to next fire is then selected probabilistically as in the SSA \cite{Gillespie1976} and the particle(s) to participate in the transformation is (are) selected randomly from the set of matches. When the rule fires, transformations are applied to the reactant complexes to create the products. Since the reactants and products are determined at runtime there is no need to enumerate all species and reactions \textit{a~priori\/} as in network-based methods. This procedure is a particle-based variant of Gillespie's algorithm \cite{Yang2008, Danos2007b} and a generalization of the ``n-fold way" of Bortz~et~al.\ \cite{Bortz1975}, which was originally developed to accelerate the simulation of Ising spin systems. An efficient, open-source implementation that is compatible with BNGL models is \mbox{NFsim}, the ``network-free simulator" \cite{Sneddon2011}. Other network-free simulation tools for rule-based models include \mbox{RuleMonkey} \cite{Colvin2010}, \mbox{DYNSTOC} \cite{Colvin2009}, \mbox{SRsim} \cite{Grunert2011}, and \mbox{KaSim} \cite{Danos2007b}.
\textcolor{black}{
A recent paper \cite{Yang2011} compares the rejection-based sampling technique \cite{Yang2008} used in \mbox{NFsim} with the rejection-free approach employed in \mbox{RuleMonkey}. For models of multivalent ligand-receptor binding, rejection-based sampling was shown to be more efficient in the vicinity of the solution-gel phase boundary, while rejection-free sampling was more efficient for simulating the dynamics within the gel phase.
}

Since only the current set of molecular complexes and the transformations that can be applied to them are tracked, network-free methods can efficiently simulate systems that are intractable to network-based methods \cite{Danos2007b, Yang2008, Sneddon2011, Yang2011}. However, the explicit representation of every molecule in the system is a major shortcoming of the approach. As such, network-free methods can require large amounts of computational memory for systems that contain large numbers of particles, a potential barrier to simulating systems such as the regulatory networks of a whole cell \cite{Tomita2001, Karr2012}. A typical eukaryotic cell, for example, contains  on the order of $10^3$--$10^4$ protein-coding genes, $10^4$--$10^5$ mRNA molecules, and $10^9$--$10^{10}$ protein molecules \cite{Alon2006, Moran2010}, along with much larger numbers of metabolites, lipids, and other small molecules. Simulating a cell at this level of detail using a network-free approach would be impractical. There is a need, therefore, for new approaches that can reduce the memory requirements of network-free simulation methods.

\subsection*{Computational complexity}

A common measure of the computational cost of an algorithm is its \emph{computational complexity\/}. In basic terms, computational complexity measures how the computational cost increases as an algorithm is applied to increasingly larger data sets \cite{Cormen2001}. For the simulation methods considered in this paper, two types of computational complexity are important: (i) \emph{space complexity\/}, the number of memory units consumed during the execution of an algorithm; (ii) \emph{time complexity\/}, the number of computational steps required to complete an algorithm.

Network-based exact-stochastic simulation methods, like Gillespie's SSA \cite{Gillespie1976, Gillespie1977, Gillespie2007}, treat species as lumped variables with a population counter. Therefore, their space complexity is constant in the number of particles in the system. However, representing the reaction network has a space complexity that is linear (or worse if a reaction dependency graph is used \cite{Gibson2000, Cao2004}) in the number of reactions. Network-based SSA methods are thus space efficient for systems with large numbers of particles, but less so for systems with large numbers of reactions. The time complexity of SSA methods is more difficult to quantify. It depends on model-specific factors such as the number of reactions in the network and the values of rate constants and species concentrations, as well as methodological factors such as how the next reaction to fire in the system is selected \cite{Gillespie1976, Gibson2000, Fricke1995, Slepoy2008, Schulze2008, Cao2004, McCollum2006} and how reaction propensities are updated after each reaction firing \cite{Gibson2000, Cao2004}. However, for our purposes, what matters is that the time cost \emph{per event\/} (reaction firing) for these methods is constant in the number of particles in the system and increases with the number of reactions in the network.

Network-free methods, in contrast, represent each particle individually. Thus, their space complexity is \emph{linear\/} in the number of particles. This is the primary shortcoming of these methods, as it limits the size of system that can be feasibly simulated. However, since reactions are not enumerated, their space complexity is linear in the number of \emph{rules\/}, rather than the number of reactions. This is a key advantage for models where very large reaction networks are encoded by a small number of rules. Network-free methods also have an advantage over network-based methods in that their time complexity per event also scales with the number of rules, rather than the number of reactions. Since the number of rules in a rule-based model is typically far less than the number of reactions, this can be a substantial improvement. For example, NFsim has been demonstrated to significantly outperform network-based SSA methods for a family of Fc$\epsilon$\/ receptor signaling models with large reaction networks \cite{Sneddon2011}. We also note that for many models network-free methods have a time cost per event that is constant in the number of particles. However, for systems in which large aggregates form (e.g., models with polymerization dynamics \cite{Monine2010, Roland2008}) the cost can be significantly higher, scaling with the number of particles \cite{Danos2007b, Sneddon2011}. Nevertheless, network-free methods are still usually the best option in these cases because these types of models tend to encode very large reaction networks \cite{Sneddon2011}. 

In Table~\ref{table:complexity}, we summarize the space and time complexities for different network-based SSA variants and for the network-free algorithm. Of most relevance to the current work are the entries that show: (i) the space complexity of network-based methods is constant in the number of particles and linear (or worse) in the reaction network size; (ii) the space complexity of network-free methods is linear in the number of particles and independent of the reaction network size, depending instead on the number of rules; (iii) the time complexity of network-based methods depends on the number of reactions in the network while for network-free methods it depends on the number of rules. Network-based methods are thus the best choice for systems with large numbers of particles and a small to moderate reaction network, and network-free methods are the best choice for systems with a large reaction network and small to moderate numbers of particles. However, neither method is optimal for systems that contain \emph{both\/} a large number of particles and a large reaction network.

\begin{table*}
\centering
\caption{\textbf{Space and time complexities for network-based (SSA) and network-free (NF) stochastic simulation algorithms.} Scalings are shown with respect to particle number, $P$\/, and number of reactions, $R$\/, or rules, $\widetilde{R}$\/. For combinatorially-complex models, $\widetilde{R} \ll R$\/. Note that time complexity is given on a ``per event" (reaction/rule firing) basis. If a reaction dependency graph \cite{Gibson2000} is used, the space and time complexities of SSA methods with respect to $R$\/ depend on $d$\/, the maximum number of reactions updated after each reaction firing \cite{Gibson2000, Cao2004}. In combinatorially-complex models, $d$\/ often increases with $R$\/ (see Figure~S1 of the supporting information). The time complexity of SSA methods with respect to $R$\/ also depends on the method used for selecting the next reaction to fire in the system. Scalings are shown for three different SSA variants that use different selection methods \cite{Gillespie1976, Gibson2000, Fricke1995, Schulze2008, Slepoy2008}. Also note that optimized variants of the direct method \cite{Cao2004, McCollum2006, Fricke1995} have been shown to outperform methods with lower asymptotic complexity in some cases \cite{Cao2004}. Space and time complexities of the NF algorithm with respect to $\widetilde{R}$\/ assume no dependency graph and that the next rule to fire is selected as in Gillespie's direct method \cite{Gillespie1976}, although in principle other variants are possible.
}
\begin{minipage}{\textwidth}
\centering
\begin{tabular*}{\textwidth}{@{\hspace{20pt}}c@{\hspace{20pt}}|@{\hspace{20pt}}cc@{\hspace{20pt}}|@{\hspace{20pt}}cc}
	\multicolumn{1}{c}{} & \multicolumn{2}{l}{\makebox[2.25in]{\textbf{SSA}}} &\multicolumn{2}{l}{\makebox[2.1in]{\textbf{NF}}} \\
	& \textit{Particles} ($P$\/) & \textit{Reactions} ($R$\/) & \textit{Particles} ($P$\/) & \textit{Rules} ($\widetilde{R}$\/) \\[2pt] \hline\hline
	\textit{Space} \rule[-6pt]{0pt}{21pt} & $\mathit{O}(1)$ & $\mathit{O}(R)$\footnote{No dependency graph}, $\mathit{O}(d{\cdot}R)$\footnote{Dependency graph \cite{Gibson2000, Cao2004}} & $\mathit{O}(P)$ & $\mathit{O}(\widetilde{R})^\mathit{a}$ \\ 
	\textit{Time (per event)} \rule[-6pt]{0pt}{21pt} & $\mathit{O}(1)$ & $\mathit{O}(d)$\footnote{Logarithmic classes (with dependency graph) \cite{Fricke1995, Schulze2008, Slepoy2008}}, $\mathit{O}(d \log_2 R)$\footnote{Next-reaction method (with dependency graph) \cite{Gibson2000}}, $\mathit{O}(R)$\footnote{Direct method (with or without dependency graph) \cite{Gillespie1976}} & $\mathit{O}(1)$, $\mathit{O}(P)$\footnote{Polymerizing systems in gel phase\cite{Yang2008, Monine2010} (see Fig.~\ref{fig:tlbr}B)} & $\mathit{O}(\widetilde{R})$\footnote{Direct method-like implementation} \\ \hline\hline
\end{tabular*}
\end{minipage}
\label{table:complexity}
\end{table*}

\subsection*{Combining network-based and network-free methodologies}

The key idea pursued in this work is that memory consumption can be reduced in network-free simulators if simple species and small molecular complexes that exist in the system in large numbers are treated as population variables with counters rather than as particles. However, retaining the ability to address combinatorial complexity requires retaining the particle representation for species and complexes that are comprised of many molecules and/or have a large number of internal states. Here, we present an approach, termed the hybrid particle/population (HPP) simulation method, that accomplishes this. Given a user-defined set of species to treat as population variables, the HPP method partially expands the network around these population species and then simulates the partially-expanded model using a population-adapted particle-based method. By treating complex species as structured particles, HPP capitalizes on the reduced time complexity with respect to network size characteristic of the network-free approach. However, for the subset of species treated as population variables, we take advantage of the constant memory requirements of the network-based methodology. It is important to emphasize that in the HPP approach it is the \emph{system\/} that is represented in a hybrid manner, as a collection of particles and population variables. The underlying simulator remains the same particle-based variant of Gillespie's algorithm that is used in existing network-free simulators \cite{Yang2008, Danos2007b}, but with small modifications to support population variables. This distinguishes HPP from other types of hybrid methods that combine different simulation methodologies, e.g., ODE/SSA integrators~\cite{Haseltine2002, Kiehl2004, Takahashi2004, Vasudeva2004, Burrage2004, Salis2005, Salis2006, Griffith2006, Wylie2006, Liu2012}.
%

\subsection*{Related work}

\textcolor{black}{
While numerous rule-based modeling frameworks have been developed, little has been done with regard to hybrid particle/population simulation. Kappa \cite{Danos2004, Danos2007a} has the concept of ``tokens," which are structureless population-type species. Modelers can write hybrid models in terms of both structured ``agents" and structureless tokens and simulate them using KaSim~3, the most recent version of the Kappa-compatible network-free simulator (\url{https://github.com/jkrivine/KaSim}). However, there is no facility for transforming a model written exclusively in terms of agents into a hybrid form, as in our HPP method. Bittig~et~al.\ \cite{Bittig2011} have developed a spatial rule-based language called ML-Space that builds upon the multi-level language ML-Rules \cite{Maus2011}. ``Entities" that are assigned optional attributes such as shape, volume, and position in continuous space are automatically treated as particles diffusing via Brownian motion, while those without these attributes are treated as population variables reacting and diffusing within a discretized space (subvolumes). For non-spatial models, the population-based network-free algorithm (PNFA) of Liu~et~al.\ \cite{Liu2010} employs a similar philosophy: all multi-state (structured) species are automatically treated as particles, while single-state species are treated as population variables. Both ML-Space and PNFA lack a general representation of intermolecular bonding, which makes it difficult to account for combinatorial complexity associated with aggregation processes \cite{Hlavacek2003, Yang2011}. Falkenberg~et~al.\ \cite{Falkenberg2013} have proposed a hybrid deterministic/stochastic method that specifically addresses the problem of aggregation. Their approach first calculates occupancy probabilities as a function of time for all binding-site types by treating them as population variables and numerically integrating an associated set of deterministic ODEs describing the binding/unbinding kinetics. An ensemble of system states is then obtained by randomly distributing bonds, based on these probabilities, among a finite number of discrete molecules. The method assumes that inter- and intra-molecular bond formations occur with equal rates. Thus, although efficient for problems with high symmetry, its applicability to more general cases may be limited.%
}

\textcolor{black}{
Other approaches aimed at improving the efficiency of rule-based simulations include ``on-the-fly" network generation \cite{Lok2005, Faeder2005, Blinov2005}, where the reaction network is gradually built up by adding reactions only when new species appear in the system. The approach has only been developed within the context of discrete-stochastic simulation and has been shown to be significantly less efficient than network-free approaches when applied to combinatorially-complex models \cite{Yang2008, Creamer2012}. An alternative approach to reducing computational cost is exact model reduction (EMR) \cite{Borisov2005, Borisov2006, Conzelmann2008, Borisov2008, Feret2009, Danos2010}. EMR aims to reduce the state space of a rule-based model while preserving the exact system dynamics with respect to observable quantities. These methods can achieve dramatic reductions in model complexity when applied within the context of ODEs, so long as the model does not contain significant cooperative or allosteric interactions \cite{Borisov2008, Danos2010}. EMR for stochastic simulations, however, has so far been less successful (see \url{http://infoscience.epfl.ch/record/142570/files/} \url{stochastic_fragments.pdf}). 
}

\section*{Methods}

\subsection*{Example models}

We have tested the performance of the HPP method by applying it to four example models, summarized in Table~\ref{table:models} and discussed in further detail below. All of the models are biologically relevant and are either taken directly from the literature or are based on models taken from the literature. Complete BNGL encodings, HPP configuration files (containing actions for loading models, defining population maps, and executing simulations), and partially-expanded versions of all example models are provided as Texts~S5--S17 of the supporting information.

\begin{table*}
\centering
\caption{\textbf{Summary of example models used to test the performance of the HPP method.} Number of particles is for an \mbox{NFsim} simulation of a full cell volume ($f\!=\!1$). Fractional cell volumes as low as 0.001 and as high as 1 are used in the performance analyses (see ``Example models" for details). Number of rules after PNE includes the population-mapping rules (one per population species).}
\renewcommand{\arraystretch}{1.5}
\begin{tabular*}{\textwidth}{l@{\hspace{20pt}}c@{\hspace{20pt}}c@{\hspace{20pt}}c@{\hspace{20pt}}c@{\hspace{20pt}}c@{\hspace{20pt}}c@{\hspace{20pt}}c}
			& & & & \textbf{Particles} & \textbf{Population} & \textbf{Rules after} & \\[-6pt]
		\multicolumn{1}{c}{\textbf{Model}} & \textbf{Rules} & \textbf{Reactions} & \textbf{Species} & \textbf{({\boldmath$f\!=\!1$})} & \textbf{species} & \textbf{PNE} & \textbf{\texttt{t\_end} (s)} \\ \hline\hline
	\textbf{TLBR} \cite{Monine2010, Sneddon2011, Goldstein1984} 
		& 4 & $\infty$ & $\infty$ & $5.3\!\times\!10^6$ & 2 & 9 & 500 \\
	\textbf{Actin} \cite{Sneddon2011, Roland2008} 
		& 21 & $\infty$ & $\infty$ & $1.2\!\times\!10^6$ & 2 & 25 & 1000 \\
	\textbf{Fc{\boldmath$\epsilon$}\/RI} \cite{Sneddon2011, Faeder2003, Goldstein2004} 
		& 24 & 58\,276 & 3744 & $6.9\!\times\!10^6$ & 1 / 6 & 25 / 38 & 2400 \\
	\textbf{EGFR} \cite{Blinov2006a, Stites2007, Fujioka2006} 
		& 113 & 415\,858 & 18\,950 & $2.2\!\times\!10^6$ & 29 & 159 & 1200\\\hline\hline
\end{tabular*}
\label{table:models}
\end{table*}

\subsubsection*{Trivalent-ligand bivalent-receptor}

The trivalent-ligand bivalent-receptor (TLBR) model is a simplified representation of receptor aggregation following multivalent ligand binding. TLBR has biological relevance to antigen-antibody interaction at the cell surface, where bivalent IgE-Fc$\epsilon$\/RI receptor complexes aggregate in the presence of multivalent antigen \cite{Goldstein1984}. A theoretical study of the TLBR system was presented by Goldstein and Perelson \cite{Goldstein1984}, who derived analytical conditions  for a solution-gel phase transition in terms of binding equilibrium constants, free ligand concentration, and receptors per cell. A more recent study considered the effects of steric constraints and ring closure on the solution-gel phase transition \cite{Monine2010}.

Despite its simplicity, the TLBR system experiences a state-space explosion near the solution-gel phase boundary. A computational study by Sneddon~et~al.\ using \mbox{NFsim} \cite{Sneddon2011} reproduced the analytical results of Goldstein and Perelson. Due to large excesses of ligand and receptor under certain conditions, TLBR is a natural test case for HPP. We simulated the TLBR system using HPP with free ligand and receptor treated as population species. All simulations were performed with parameters as defined in Monine~et~al.\ \cite{Monine2010}, which lie within the solution-gel phase coexistence region. A cell-scale simulation assumed 1~nl extracellular volume per cell ($10^6$ cells/ml) with 8.3~nM ligand and $3\!\times\!10^5$ receptors per cell. Simulations were performed at fractional cell volumes, $f$\/, ranging from 0.001 to 0.1 with a lumping rate constant $\mathtt{k\_lump}\!=\!10\,000$/s (see below).

\subsubsection*{Actin polymerization}

Actin polymerization plays a key role in cell morphology and motility \cite{Pollard2003, Lacayo2007}.  Roland~et~al.\ \cite{Roland2008} presented a dynamic model of actin polymerization featuring filament elongation by monomer addition, stabilization by ATP hydrolysis, and severing mediated by actin depolymerizing factor (ADF)/cofilin. Sneddon~et~al.\ \cite{Sneddon2011} presented a rule-based formulation of the Roland~et~al.\ model and replicated their results using \mbox{NFsim}. The model features an excess of actin monomer and ADF molecules. Therefore, we speculated that substantial memory reduction would be possible using the hybrid approach. We applied HPP to the Sneddon~et~al.\ rule-based model of actin dynamics (hereafter referred to as the Actin model) with actin monomer and ADF treated as population species. A cell-scale simulation assumed 1~pl intracellular volume with 1~$\mu$\/M actin monomer and 1~$\mu$\/M ADF/cofilin. Simulations were performed at fractional cell volumes, $f$\/, ranging from 0.01 to 1 with a lumping rate constant $\mathtt{k\_lump}\!=\!10\,000$/s.

\subsubsection*{Fc$\epsilon$\/RI signaling}

Fc$\epsilon$\/RI is a membrane receptor that binds IgE antibodies. Signaling through Fc$\epsilon$\/RI regulates basophilic histamine release in response to IgE antibody-antigen interaction \cite{Stone2010}. Faeder~et~al.\ \cite{Goldstein2002, Faeder2003} developed a rule-based model of Fc$\epsilon$\/RI receptor assembly and activation in which receptor dimerization/clustering is mediated by chemically cross-linked IgE, which serve as multivalent ligands. Dimerized receptors are  transphosphorylated, leading to Syk and Lyn recruitment and phosphorylation. Sneddon~et~al.\ \cite{Sneddon2011} presented several extensions of the Faeder~et~al.\ model, including the \emph{gamma2\/} variant with two $\gamma$\/ phosphorylation sites. Particle-based \mbox{NFsim} simulations of the \emph{gamma2\/} model were found to be substantially faster than network-based SSA simulations. 

Due to the excess of free ligand, the HPP method was applied to the \emph{gamma2\/} model to reduce memory consumption. The method was applied with two different sets of population species. In the first case, only free ligand was treated as a population species (Fc$\epsilon$\/RI:1). In the second, cytosolic Lyn and all four phosphorylation states of cytosolic Syk were also treated as populations (Fc$\epsilon$\/RI:6). A cell-scale simulation assumed 1~pl intracellular volume with 1~nl extracellular space per cell ($10^6$ cells/ml), 10~nM ligand, and $4\!\times\!10^5$ receptors per cell. Simulations were performed at fractional cell volumes, $f$\/, ranging from 0.001 to 0.1 with a lumping rate constant $\mathtt{k\_lump}\!=\!10\,000$/s.

\subsubsection*{EGFR signaling}

A model of signaling through the epidermal growth factor receptor (EGFR), beginning with ligand binding and concluding with nuclear phospho-ERK activity, was constructed by combining three existing models: (i) a rule-based model of EGFR complex assembly \cite{Blinov2006a}; (ii) a Ras activation model \cite{Stites2007}; (iii) a pathway model of Raf, MEK and ERK activation \cite{Fujioka2006}. Ras activation was coupled to the EGFR complex assembly by treating receptor-recruited Sos as the Ras GEF. Activated Ras was coupled to the Raf/MEK/ERK cascade through RasGTP-Raf binding and subsequent phosphorylation of Raf. Parameters for the combined model were obtained from the respective models. However, parameters governing Ras-GEF (i.e., Sos) activity had to be changed from their original values \cite{Stites2007} in order to account for the known GEF-mediated activation of Ras \cite{Overbeck1995}. Specifically, we used $K_{M,\mathrm{GDP}}\!=\!K_{M,\mathrm{GTP}}\!=\!1.56\!\times\!10^{-7}$~M and $D\!=\!1000$ (unitless). 

Free EGF and Raf-, MEK-, and ERK-based species were treated as population species in the hybrid variant. Ras-based species were also treated as populations except for those that include a Sos molecule. A cell-scale simulation assumed 0.94~pl cytosolic and 0.22~pl nuclear volume, with 0.94~pl extracellular space, 10~nM ligand, and $4\!\times\!10^5$ receptors per cell. Simulations were performed at fractional cell volumes, $f$\/, ranging from 0.01 to 1 with a lumping rate constant $\mathtt{k\_lump}\!=\!100\,000$/s.

\subsection*{Performance metrics}

HPP was evaluated for peak memory use, CPU run time, and accuracy as compared to particle-based \mbox{NFsim} simulations.  For models where network generation is possible (Fc$\epsilon$\/RI and EGFR), comparisons were also made to SSA simulations (as implemented within \mbox{BioNetGen} \cite{Faeder2009}).  All simulations were run on a 2 $\times$\/ Intel Xeon E5520 @ 2.27~GHz (8 cores, 16 threads, x86\_64 instruction set) with 74~GB of RAM running the GNU/Linux operating system. To ensure that each process had access to 100\% of the compute cycles of a thread, no more than 12 simulations were run simultaneously.

\subsubsection*{Peak memory}

Average peak memory usage for each simulation method was calculated based on seven independent simulation runs. Peak memory for each run was evaluated by peak virtual memory allocation reported by the operating system with the command ``\texttt{cat /proc/<PID>/status}". For all tested models, peak memory was achieved early in the simulation and remained steady throughout (data not shown).

\subsubsection*{CPU run time}

Average CPU run time for each simulation method was calculated based on seven independent simulation runs using clock time as a metric. Clock time for each run was recorded using the \texttt{Time::HiRes} Perl module. Run time included initialization as well as the simulation phase. Partial network expansion for HPP simulations was a one time cost, typically a few seconds, and was not included in the calculation.

\subsubsection*{Accuracy}

Simulation accuracy was quantified using several approaches. First, since HPP, NFsim, and SSA are all exact-stochastic methods, they should all produce statistically the same number of reaction firings. To verify this, for all tested models the total number of reaction firings was recorded for each of 40 independent simulation runs of each method (firings of population-mapping rules were subtracted from the total in HPP simulations). The Mann-Whitney~U test \cite{Wilcoxon1945, Mann1947} was then used to test the null hypothesis that none of the methods produces a larger number of reaction firings.

For the TLBR and Actin models, we further compared equilibrium distributions for key observables. These include the number of receptor clusters in the TLBR model and the length of actin polymers in the Actin model. 10\,000 samples were collected over 100\,000 seconds of simulated time and distributions were compared by binning samples (20 bins) and performing a two-sample chi-squared test \cite{Pearson1900}. For the Fc$\epsilon$\/RI and EGFR models, we compared dynamic trajectories for key observables. These include $\gamma$-phosphorylated receptor and receptor-recruited, $\alpha$-phosphorylated Syk in the Fc$\epsilon$\/RI model, and activated Sos and nuclear phosphorylated ERK in the EGFR model. Due to complications of autocorrelation, a statistical test was not applied to the dynamic trajectory comparison. Instead, moving averages and 5--95\% frequency envelopes, based on 40 simulation runs of each method using a sampling window of 10~s, were plotted for inspection by eye.

\subsection*{Software}%
\label{sec:software}

All HPP and NFsim simulations reported in this work were run using \mbox{NFsim} version 1.11, which is available for download at \url{http://emonet.biology.yale.edu/nfsim}. All simulations (SSA included) were invoked through \mbox{BioNetGen} version 2.2.4, which implements the hybrid model generator and is distributed with \mbox{NFsim}~1.11. Instructions for running simulations with \mbox{BioNetGen} (ODE, SSA, and HPP) can be found in Secs.~S3.2 and S3.3 of Text S1 and Refs.~\cite{Faeder2009, Sekar2012}. \mbox{NFsim} and \mbox{BioNetGen} source code are available at \url{http://code.google.com/p/nfsim} and \url{http://code.google.com/p/bionetgen}, respectively. Additional documentation for \mbox{BioNetGen} can be found at \url{http://bionetgen.org}. 

\section*{Results}

\subsection*{A hybrid particle/population simulation approach}

\begin{figure}[!t]
\caption{\textbf{Basic workflow of the HPP simulation method.} Given a rule-based model and a user-specified set of population-mapping rules (which define the population species), partial network expansion (PNE) is performed to generate a hybrid version of the original model. The hybrid model is then passed to a population-adapted network-free simulator (e.g., \mbox{NFsim}~1.11), which generates the time-evolution trajectories for all observable quantities specified in the original model.}
\label{fig:hppWorkflow}
\begin{center}
\includegraphics[width=0.4\textwidth]{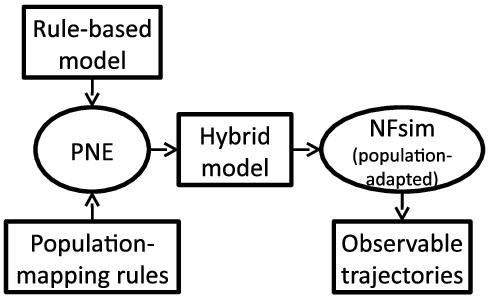}
\end{center}
\end{figure}

In this section, we first present an approach, termed ``partial network expansion," for transforming a rule-based model into a dynamically-equivalent, partially-expanded form. We then describe a simple modification to the network-free simulation protocol that permits simulation of the transformed model as a collection of both particles and population variables. We refer to the combination of these methods as the hybrid particle/population (HPP) simulation method. The basic workflow is shown in Fig.~\ref{fig:hppWorkflow}. 

The HPP approach is analogous to the coupled procedure of network generation and simulation described above, where a rule-based model is first transformed into a \emph{fully-expanded\/} reaction network and then simulated as a collection of population variables (i.e., species) using a network-based simulator. The obvious differences are that in HPP the network is only partially expanded and the system can only be simulated stochastically using a population-adapted network-free simulator. The partial network expansion algorithm has been implemented within the open-source rule-based modeling package \mbox{BioNetGen} \cite{Blinov2004, Faeder2009, Sekar2012} and resulting hybrid models can be simulated using version 1.11 (or later) of the network-free simulator \mbox{NFsim} \cite{Sneddon2011}, which has been modified to handle population-type species. For convenience, we adhere in this paper to the BNGL syntax, which is summarized in Sec.~S3.1 of Text~S1 of the supporting material. However, the HPP method is generally applicable to any rule-based modeling language for which there exists a network-free simulator capable of handling a mixed particle/population system representation, e.g., KaSim~3.x for Kappa language models (see \url{https://github.com/jkrivine/KaSim}).

\subsubsection*{Population species and population-mapping rules}

Given a rule-based model, the first step in the HPP approach is to select a subset of species to treat as ``lumped" population variables. There are no hard-and-fast rules for doing this but, generally speaking, species that are good candidates for a population treatment (i) have a small number of components and internal states, (ii) participate in a small number of rules, and (iii) maintain a large population throughout the course of a simulation. An example is a simple ligand species that exists in great excess in the extracellular environment and interacts with cell surface receptors. It is our experience that these simple rules of thumb, combined with the experience and intuition of the modeler, are usually sufficient for selecting an adequate set of population species. However, in some cases a more systematic approach may be desirable. We will return to this topic below.

For now, however, let us assume that we have selected a suitable set of population species. The next step in the HPP approach is to map each of these to an associated \emph{unstructured\/} species. The mapping is accomplished by defining a \emph{population-mapping rule\/}, which follows the same syntactic conventions as a standard BNGL rule. For example, the rule
\begin{center}
\texttt{Egf(r) -> pop\_Egf()  k\_lump}
\end{center}
maps the unbound EGF ligand, \texttt{Egf(r)}, to the unstructured species \texttt{pop\_Egf()}. To avoid confusion, we will henceforth refer to species on the reactant side of a population-mapping rule, such as \texttt{Egf(r)}, as \textit{structured population species\/} and to those on the product side as \textit{unstructured population species\/}. Importantly, unstructured population species differ from conventional unstructured molecules in BNGL in that they possess a property, called a \emph{count\/}, which records their current population (see Sec.~S3.3 of Text~S1 and Texts~S4, S7, S10, S13, S14, and S17 to see how the \texttt{population} keyword is used to make this distinction). The action of the population-mapping rule above is thus to delete the \texttt{Egf(r)} molecule and to \emph{increment by one\/} the count of \texttt{pop\_Egf()}. The role of the rate parameter \texttt{k\_lump}, termed the \emph{lumping rate constant\/}, will be explained in detail below.

\subsubsection*{Partial network expansion}

Ultimately, our goal in the HPP method is to replace in the simulation environment large numbers of indistinguishable particles with small numbers of lumped objects containing population counters (the unstructured population species), thus significantly reducing memory usage. In order to accomplish this without losing any information regarding the dynamics of the system, we must partially expand the rule set of the original model until all interactions and transformations in which the structured population species participate \emph{as reactants\/} (see below) are enumerated. We can then swap the structured species with their unstructured counterparts, which have been specified via the population-mapping rules. We refer to this procedure as partial network expansion (PNE).

The PNE algorithm is comprised of three basic steps, which are applied to each rule of a rule-based model:
\begin{enumerate}
	\item{For each reactant pattern in the rule, identify all matches of that pattern into the set of structured population species. Also collect a self-match of the reactant pattern \emph{unless it equals\/} one of the population species (this can only happen if the reactant pattern is a fully-specified species; see below for further discussion).}
	\item{Derive an expanded set of rules by applying the rule to all possible combinations (the cartesian product) of the  pattern matches collected in Step~1.}
	\item{For each derived rule from Step~2, replace each instance of a structured population species with its unstructured population counterpart.}
\end{enumerate}
The result is an expanded rule set consisting of three general types of rules: (i) particle rules, in which all reactants are conventional reactant patterns; (ii) mixed particle/population rules, where at least one reactant is a conventional reactant pattern and one is an unstructured population species; (iii) pure population \emph{reactions\/}, where all reactants are unstructured population species. This expanded rule set has the property that every possible action of the original rule set on the population species is enumerated while actions on particle objects remain pattern-based (i.e., non-enumerated). For a more formal presentation of the PNE algorithm, complete with pseudocode, we direct the reader to Sec.~S4.2 of Text~S1.

\subsubsection*{Role of the population-mapping rules}

After completion of PNE, the final step in transforming a rule-based model into a form that can be simulated as a hybrid particle/population system is to append the population-mapping rules to the expanded rule set. The reason for doing this is not immediately obvious. We have seen above that the population-mapping rules specify which structured species are to be replaced in the transformed model with population variables. However, an obvious question to ask is why we have chosen to specify this information via a set of reaction rules, rather than simply as a list of species to be lumped. The answer is combinatorial complexity. 

\begin{figure}[!t]
\caption{
\textbf{Simple illustration of ambiguity in the products of reaction rules.} (A) A simple rule encodes the reversible binding of two molecule types, \texttt{A} and \texttt{B}. (B)--(D) If both molecules have multiple binding sites then they may be present within arbitrarily complex complexes. Breaking the bond between \texttt{A} and \texttt{B} thus produces a variety of product species, some of which may correspond to population species and others not. Dashed line represents a bond addition/deletion operation.
}
\label{fig:popMaps}
\begin{center}
\includegraphics[width=0.5\textwidth]{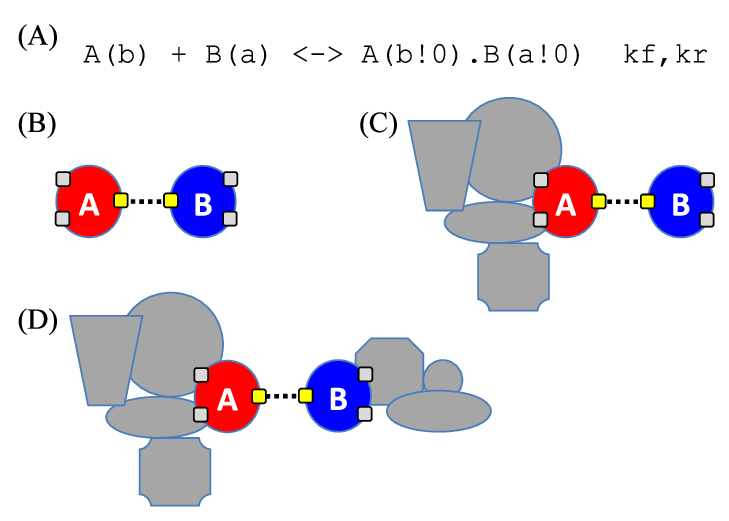}
\end{center}
\end{figure}

As explained above, systems that are combinatorially complex are comprised of a relatively small number of constituent parts but exhibit an explosion in the number of potential species and reactions due to the myriad number of ways in which these parts can be connected and arranged. Rule-based modeling is effective in representing these systems because it focuses only on the portions of molecular complexes that affect biochemical reactivity, not on entire species. However, a consequence of this approach is that there is often ambiguity regarding the products of a reaction rule. A rule may describe the breaking of a bond between two molecules, for example, but the exact composition of the resulting complexes is left necessarily ambiguous (see Fig.~\ref{fig:popMaps}).

With regard to the HPP approach, this ambiguity in the products of a reaction rule complicates the process of PNE. Application of a reaction rule to one complex may produce a population species, whereas application of the same rule to a different complex may not. Distinguishing between cases where population species are produced and where they are not is difficult, and may even be impossible if the system is combinatorially complex. Thus, the strategy that we have adopted here is to expand the network out only to the point where all population species \emph{on the reactant side\/} are enumerated and to handle the ambiguity in products by adding the population-mapping rules to the rule set.  The role of the population-mapping rules is thus to detect any instances of structured population species that appear in the simulation environment as products of a rule application and to gather them up into the unstructured population pool. 

This returns us to the issue of the lumping rate constant, \texttt{k\_lump}. In Step~1 of the PNE algorithm, if a reactant pattern equals a population species then we discard the self-match (the structured version of the population species). To see why we do this, consider the binding rule depicted in Fig.~\ref{fig:popMaps}A. However, different from Figs.~\ref{fig:popMaps}B--D, assume that molecules \texttt{A} and \texttt{B} have only \emph{one\/} binding site each. If we choose to lump the unbound molecules then we must define the following population-mapping rules:
\begin{center}
	\texttt{A(b) -> pop\_A() k\_lump},\\
	\texttt{B(a) -> pop\_B() k\_lump}.\\
\end{center}
Obviously, these structured population species are equivalent to the reactant patterns in Fig.~\ref{fig:popMaps}A. However, let us choose \emph{not\/} to discard the self-matches in this case. PNE would then generate the following four derived rules:
\begin{center}
\begin{tabular}{rcc} 
	\texttt{A(b) + B(a)} & \texttt{->} & \texttt{A(b!0).B(a!0) kf},\\ 
	\texttt{pop\_A() + B(a)} & \texttt{->} & \texttt{A(b!0).B(a!0) kf},\\
	\texttt{A(b) + pop\_B()} & \texttt{->} & \texttt{A(b!0).B(a!0) kf},\\
	\texttt{pop\_A() + pop\_B()} & \texttt{->} & \texttt{A(b!0).B(a!0) kf}.\\
\end{tabular} 
\end{center}
We see that the first three of these rules have conventional (structured) reactant patterns. However, if \texttt{k\_lump} is sufficiently large then particle instances of \texttt{A(b)} and \texttt{B(a)} will never exist in the system long enough to be matched to these patterns. Thus, these rules can be safely discarded, which is equivalent to discarding the self-match in Step~1 of the PNE algorithm. Retaining only the fourth derived rule (the pure population version) simplifies the process and keeps the size of the derived rule set to a minimum.

The consequence of this is obviously that the HPP method is formally exact \emph{only\/} for an infinite lumping rate constant. From a practical point of view, this could be a problem if the network-free simulator being used does not support infinite rates (e.g., NFsim currently does not). However, our performance tests indicate that as long as \texttt{k\_lump} is ``large" with respect to the model dynamics then essentially exact results can be obtained (see Figs.~\ref{fig:tlbr}--\ref{fig:egfr}, panels C and D). Nevertheless, we have implemented in \mbox{BioNetGen} a ``safe" mode for PNE that retains all of the self-matches and, hence, produces exact results for \emph{any\/} value of \texttt{k\_lump} (see Sec.~S3.3 of Text~S1 for instructions on how to call this method). For a select number of examples, we have confirmed that both approaches give essentially identical results for sufficiently large \texttt{k\_lump} and that the ``safe" mode is less efficient (data not shown). 

\subsubsection*{Simple example of PNE}

\begin{figure*}[!t]
\caption{\textbf{Simple receptor activation model in BNGL format.} Abridged; see \textcolor{black}{Text~S2} of the supporting material for the complete model and Text~S3 for the population-mapping rules.}
\label{fig:recAct}
\begin{center}
\includegraphics[width=0.9\textwidth]{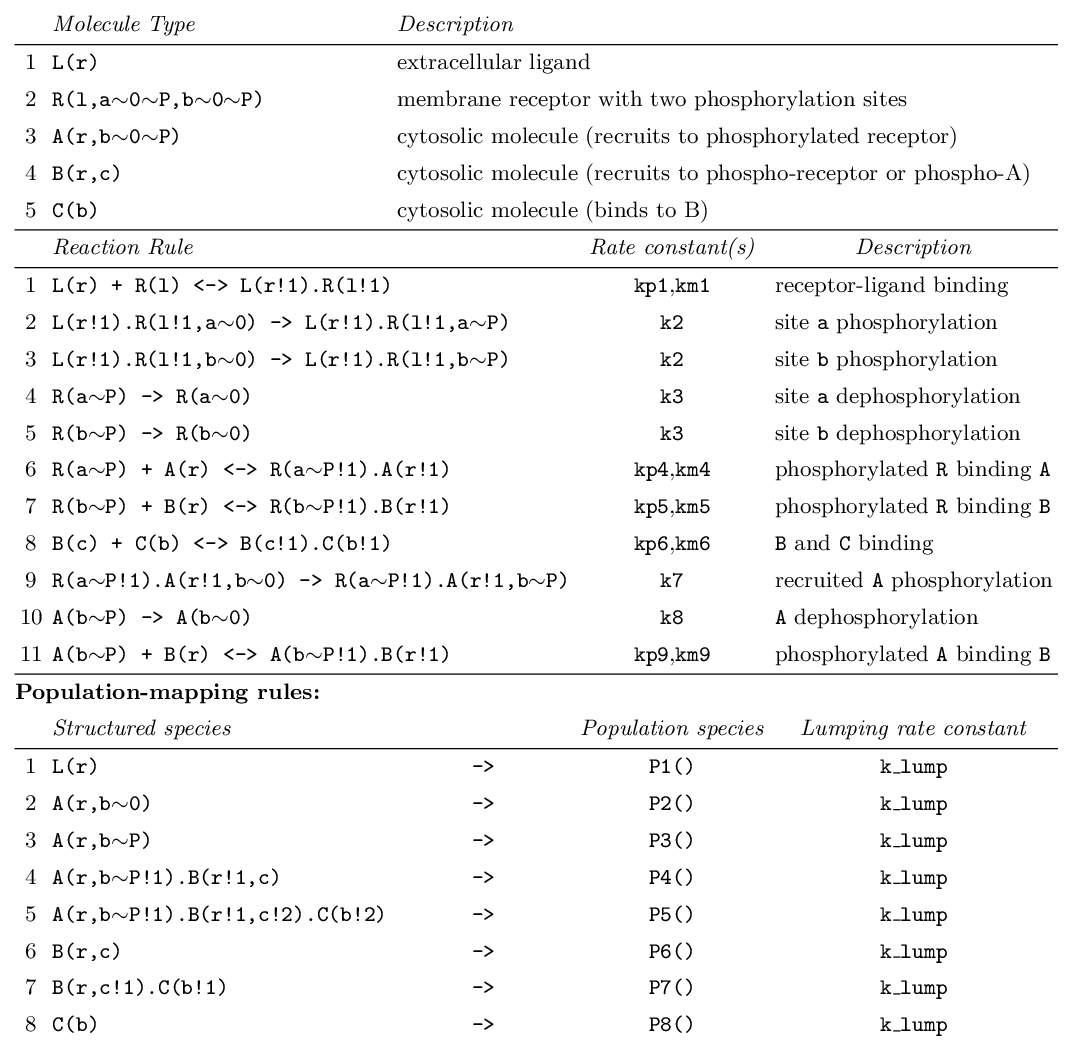}
\end{center}
\end{figure*}

PNE is best illustrated through an example. In Fig.~\ref{fig:recAct}, we present a simple rule-based model of receptor activation (for brevity, parameters, initial populations, and output observables are omitted; see \textcolor{black}{Text~S2} of the supporting material for the complete model in BNGL format). The model includes a ligand, \texttt{L}, its cognate receptor, \texttt{R}, and three cytosolic proteins, \texttt{A}, \texttt{B}, and \texttt{C}, that are recruited to the phosphorylated receptor. The 16 rules (six unidirectional and five reversible), describing ligand-receptor binding, receptor phosphorylation/dephosphorylation, and protein recruitment, encode a reaction network comprised of 56 species and 287 reactions. In applying the HPP method, eight species are selected for lumping: free ligand, free \texttt{A}, \texttt{B} and \texttt{C}, and complexes of  \texttt{A}, \texttt{B} and \texttt{C} that exclude the receptor. Receptor complexes are treated as particles because there are many possible receptor configurations (48 total).

\begin{figure*}[!t]
\caption{\textbf{Partial network expansion (PNE) applied to Rule~11f of Fig.~\ref{fig:recAct}.} See \textcolor{black}{Text~S4} of the supporting material for the complete, partially-expanded model.}
\label{fig:pneExample}
\begin{center}
\includegraphics[width=0.9\textwidth]{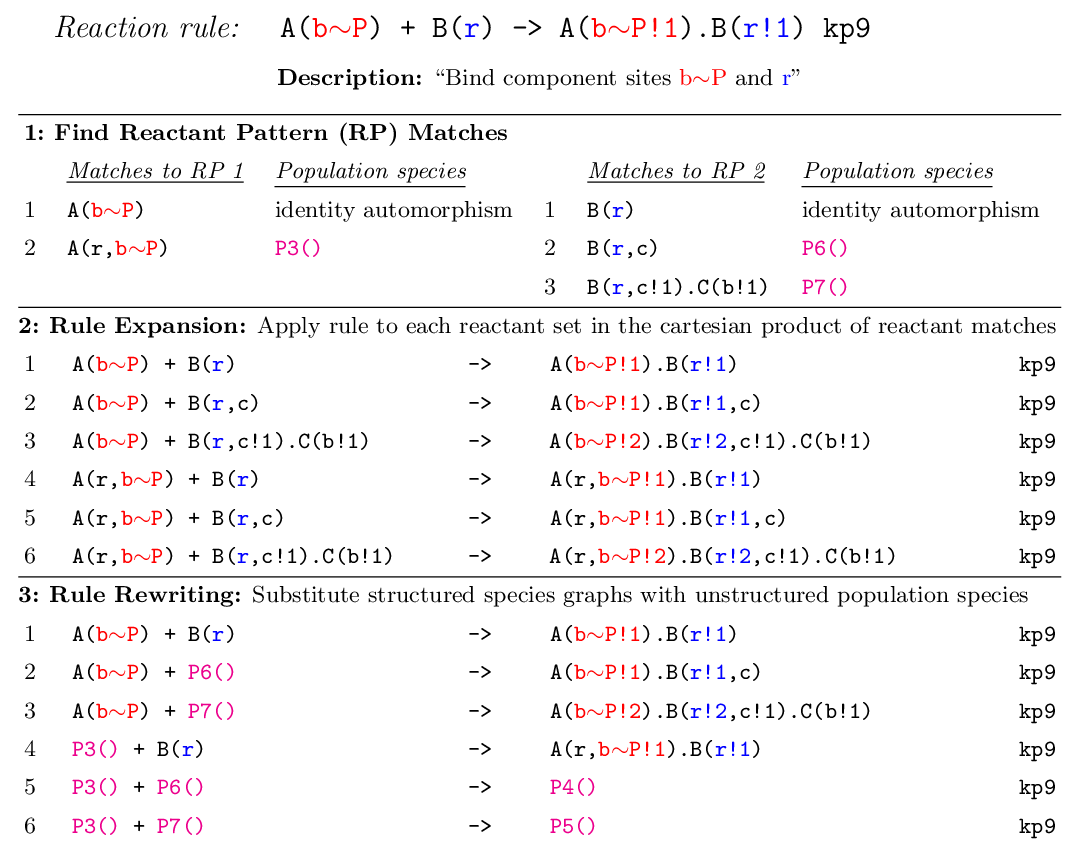}
\end{center}
\end{figure*}

In Fig.~\ref{fig:pneExample}, a step-by-step application of PNE to rule 11f (forward) of Fig.~\ref{fig:recAct} is presented. First, both reactant patterns are matched to the structured population species. Reactant pattern~1 has one match, while reactant pattern~2 has two. Note that since neither reactant pattern exactly equals a species (i.e., is isomorphic to one) the self match (identity automorphism) is added to the reactant match list in both cases. Next, the rule is applied to each possible reactant set (the cartesian product of the reactant match lists). This results in a set of six derived rules. The structured population species are then replaced in these rules by their associated unstructured species, resulting in one pure particle rule (the original rule), three mixed particle/population rules, and two pure population reactions. Including the population-mapping rules, the hybrid model contains a total of 42 rules, more than the original 16 but significantly less than the 287 reactions of the fully-expanded network. The complete partially-expanded HPP model in BNGL format can be found in Text~S4 of the supporting material.

\subsubsection*{Population-adapted network-free simulation}

Although modified relative to the original, the hybrid model generated from PNE remains properly a rule-based model. As such, it can, in principle, be simulated with any of the network-based (after network generation) and network-free simulation methods described above. However, the advantage of recasting the original model into the hybrid form is that it can be represented as a collection of particles and population objects and simulated using a modified network-free method that has the following attributes: (i) a population count property for each molecule object; (ii) a transformation that performs population increments and decrements; (iii) a method for calculating population-weighted propensities (rates). Examples of population-adapted network-free simulators are NFsim~1.11 and KaSim~3.x.

The population-weighted propensity of a rule $R_\mu$\/ can be calculated as
\begin{equation}
a_\mu = \frac{k_\mu}{s_\mu} \prod_{r=1}^{M_\mu} \left( \sum_{x=1}^X \rho(x) \eta_{\mu,r}(x) \right).
\label{eq:hybrid_aMu}
\end{equation}
Here, $k_\mu$\/ is the rate constant (more generally, the ``single-site rate law" \cite{Faeder2009}), $s_\mu$\/ is the symmetry factor (see Note~4.21 of Ref.~\cite{Faeder2009}), $M_\mu$\/ is the number of reactant patterns in the rule (i.e., the \emph{molecularity\/}), $X$\/ is the total number of complexes in the system, $\rho(x)$ is the population of complex $x$\/ (unity in the case of particles), and $\eta_{\mu,r}(x)$ is the number of matches of reactant pattern $r$\/ into complex $x$\/ (unity or zero for unstructured population species, i.e., the species either is the reactant or it is not). The difference between Eq.~\ref{eq:hybrid_aMu} and the formula used for calculating propensities in standard network-free simulators is the term $\rho(x)$; a fully particle-based network-free calculation is recovered if all $\rho(x)=1$. Conversely, the difference between Eq.~\ref{eq:hybrid_aMu} and the formula used in network-based SSA simulators is the term $\eta_{\mu,r}(x)$; a fully population-based calculation is recovered if all $\eta_{\mu,r}(x)=0$ or $1$, in which case $X$\/ is the total number of species in the network. Equation~\ref{eq:hybrid_aMu} thus generalizes the concept of propensity for hybrid systems comprised of both particles and population variables.

Also note that for symmetric population reactions, e.g., \texttt{pop\_A() + pop\_A() -> A(a!0).A(a!0)}, the possibility of a null event must be calculated in order to prevent reactions involving the same molecule. This is accomplished by rejecting the event with probability $1/\rho(x)$. Furthermore, since population species have zero components, if complex $x$\/ is a population species and $\eta_{\mu,r}(x)=1$, then  $\eta_{\mu,r}(y)=0$ for all $y\!\ne\!x$\/. This property is useful because it guarantees that a reactant pattern matches either particles or population species exclusively, never a mixture of both. Thus, once a rule has been selected to fire, the particles to participate in that rule can be selected from a uniform distribution rather than from a population-weighted distribution. 

\subsection*{Performance analyses}

\begin{figure}[!t]
\caption{\textbf{HPP performance analysis for the TLBR model.} (A) peak memory usage (\emph{left\/}:\ absolute, \emph{right\/}:\ relative to \mbox{NFsim}); (B) CPU run time (\emph{left\/}:\ absolute, \emph{right\/}:\ relative to \mbox{NFsim}); (C) number of reaction events fired during a simulation ($f\!=\!0.01$); (D) equilibrium distribution of number of clusters ($f\!=\!0.01$).}
\label{fig:tlbr}
\begin{center}
\includegraphics[width=0.4\textwidth]{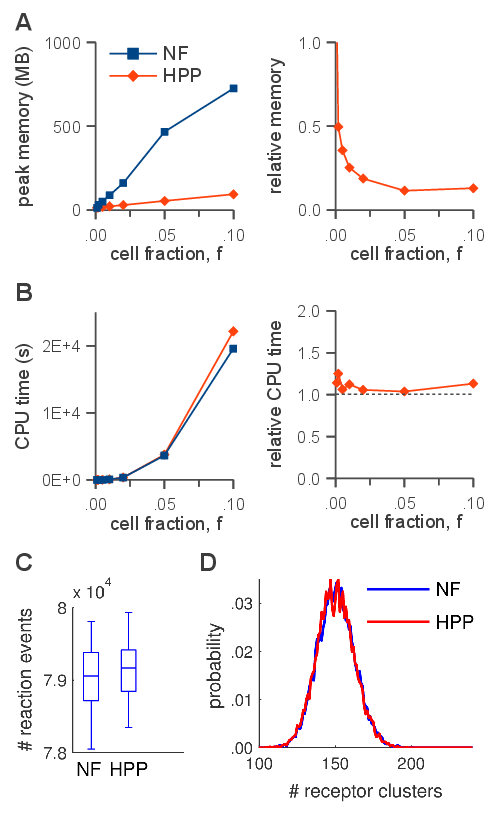}
\end{center}
\end{figure}
\begin{figure}[!t]
\caption{\textbf{HPP performance analysis for the actin polymerization model.} (A) peak memory usage (\emph{left\/}:\ absolute, \emph{right\/}:\ relative to \mbox{NFsim}); (B) CPU run time (\emph{left\/}:\ absolute, \emph{right\/}:\ relative to \mbox{NFsim}); (C) number of reaction events fired during a simulation ($f\!=\!0.01$); (D) equilibrium distribution of actin polymer lengths ($f\!=\!0.01$).}
\label{fig:actin}
\begin{center}
\includegraphics[width=0.4\textwidth]{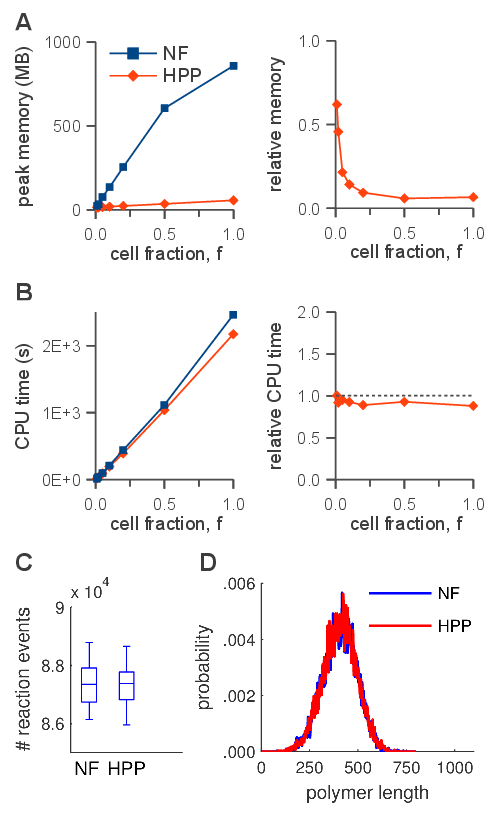}
\end{center}
\end{figure}
\begin{figure}[!t]
\caption{\textbf{HPP performance analysis for the Fc{\boldmath$\epsilon$}\/RI signaling model.} (A) peak memory usage (\emph{left\/}:\ absolute, \emph{right\/}:\ relative to \mbox{NFsim}); (B) CPU run time (\emph{left\/}:\ absolute, \emph{right\/}:\ relative to \mbox{NFsim}); (C) number of reaction events fired during a simulation ($f\!=\!0.01$); (D) timecourses (means and 5--95\% frequency envelopes; $f\!=\!0.01$) for $\gamma$-phosphorylated receptor (\emph{top\/}) and receptor-recruited, $\alpha$-phosphorylated Syk (\emph{bottom\/}). SSA timecourses are virtually indistinguishable and have been omitted for clarity.}
\label{fig:fceri} 
\begin{center}
\includegraphics[width=0.4\textwidth]{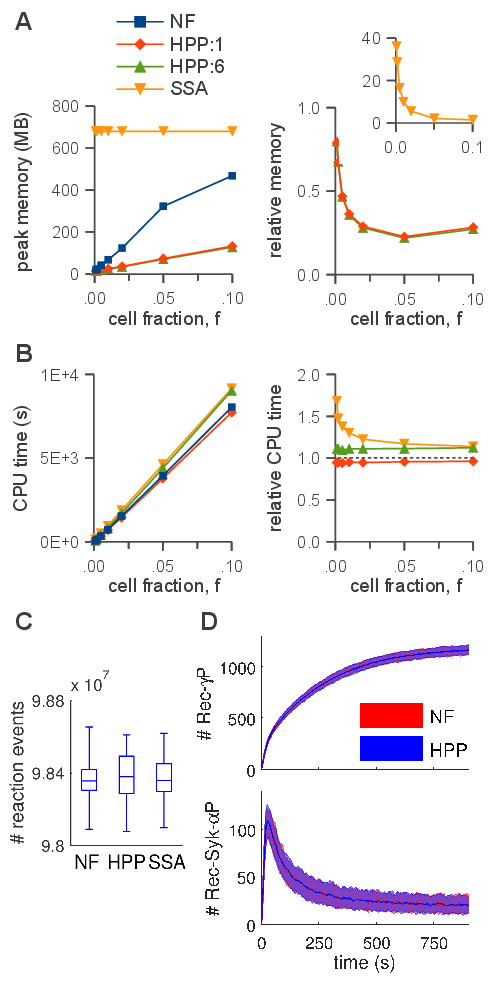}
\end{center}
\end{figure}
\begin{figure}[!t]
\caption{\textbf{HPP performance analysis for the EGFR signaling model.} (A) peak memory usage (\emph{left\/}:\ absolute, \emph{right\/}:\ relative to \mbox{NFsim}); (B) CPU run time (\emph{left\/}:\ absolute, \emph{right\/}:\ relative to \mbox{NFsim}); (C) number of reaction events fired during a simulation ($f\!=\!0.05$); (D) timecourses (means and 5--95\% frequency envelopes; $f\!=\!0.05$) for activated Sos (\emph{top\/}) and nuclear phosphorylated ERK (\emph{bottom\/}). Due to high computational expense, SSA statistics were not collected in (C) and (D).}
\label{fig:egfr}
\begin{center}
\includegraphics[width=0.4\textwidth]{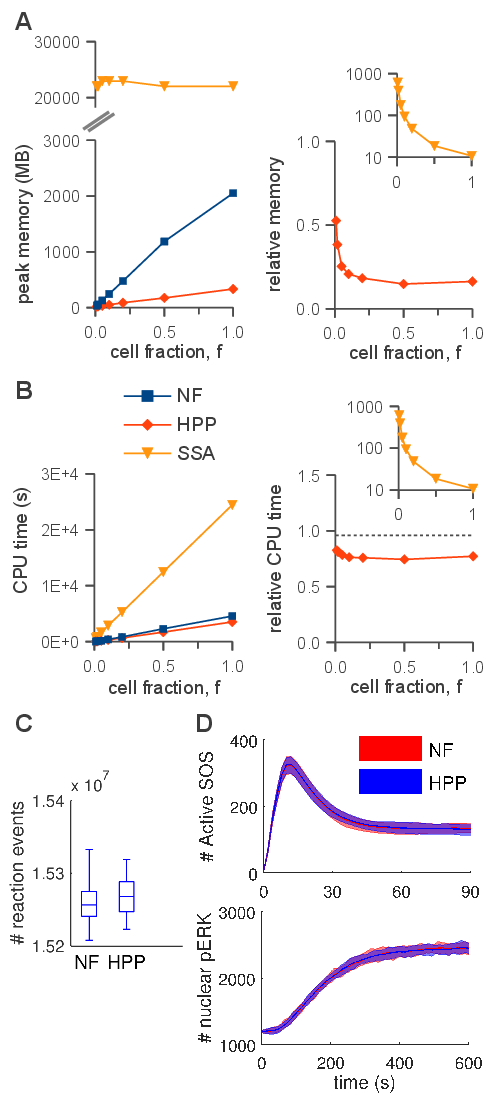}
\end{center}
\end{figure}

\subsubsection*{Peak memory use and CPU run time}

In Figs.~\ref{fig:tlbr}--\ref{fig:egfr}, panels A, we show absolute and relative (with respect to \mbox{NFsim}) peak memory use as a function of cell fraction, $f$\/, for all models considered. We see that in all tested cases HPP requires less memory than \mbox{NFsim}. For \mbox{NFsim}, we also see the expected linear relationship (Table~\ref{table:complexity}) between peak memory use and particle number (i.e., cell fraction; the slight deviation from linearity is an artifact of how memory is allocated in NFsim). For HPP, peak memory use also scales linearly with particle number, but with a smaller slope. This is the expected behavior since as the cell fraction is increased (keeping concentrations constant) a portion of the added particles, and hence memory cost, is always absorbed by the population portion of the system. Furthermore, in cases where network generation is possible (Fc$\epsilon$\/RI, Fig.~\ref{fig:fceri}A; EGFR, Fig.~\ref{fig:egfr}A), we see the expected constant relationship between memory usage and particle number for the SSA (Table~\ref{table:complexity}). We also see that the SSA requires more memory than both \mbox{NFsim} and HPP for all cell fractions considered. This is due to the high memory cost of the dependency update graph \cite{Cao2004} used in the SSA implementation within \mbox{BioNetGen}, which scales with the product of the number of reactions in the network and the number of reactions updated after each reaction firing (see Table~\ref{table:complexity}).

In Figs.~\ref{fig:tlbr}--\ref{fig:egfr}, panels B, we show absolute and relative (with respect to \mbox{NFsim}) CPU run times as a function of cell fraction. Generally speaking, HPP and \mbox{NFsim} run times are comparable in all cases, indicating that the reductions in memory use seen in Figs.~\ref{fig:tlbr}--\ref{fig:egfr}, panels A, are not achieved at the cost of increased run times. In fact, HPP is slightly faster than \mbox{NFsim} in most cases. This is because operations on population species (e.g., increment/decrement) are less costly than the graph operations applied to particles (e.g., subgraph matching). Also note in Fig.~\ref{fig:tlbr}B the expected quadratic relationship between run time and particle number for the TLBR model (Table~\ref{table:complexity}), which is due to the formation of a super aggregate near the solution-gel phase boundary \cite{Yang2008, Monine2010}.
In Figs.~\ref{fig:fceri}B and \ref{fig:egfr}B, we see that the SSA is slower than both \mbox{NFsim} and HPP for all cell fractions considered. The difference is most pronounced at small cell fractions and is much more significant for EGFR than for Fc$\epsilon$\/RI. This is expected since previous work \cite{Sneddon2011} has shown that network-free methods perform particularly well for systems with small numbers of particles and large networks (the EGFR network is significantly larger than the Fc$\epsilon$\/RI network; Table~\ref{table:models}).
Finally, we see in Fig.~\ref{fig:fceri}B that the CPU run time increases as we increase the number of species treated as populations in the Fc$\epsilon$\/RI model, even though the memory usage remains constant (Fig.~\ref{fig:fceri}A). This is interesting because it suggests that the Fc$\epsilon$\/RI:1 variant, with free ligand as the only population species, is near-optimally lumped for the cell fractions considered. We revisit the issue of optimal lumping sets below.

\subsubsection*{Accuracy}

In Figs.~\ref{fig:tlbr}--\ref{fig:egfr}, panels C, we show distributions of the number of reaction firings per simulation run for each of the simulation methods considered. It is evident that for all models the distributions, as illustrated by box plots, are similar for \mbox{NFsim}, HPP, and SSA (the latter for Fc$\epsilon$\/RI only; Fig.~\ref{fig:fceri}C). Statistically speaking, the two-sided Mann-Whitney~U test \cite{Wilcoxon1945, Mann1947} was unable to reject the null hypothesis in all cases at the 5\% significance level (TLBR:\ $p\!=\!0.25$; Actin:\ $p\!=\!0.90$; Fc$\epsilon$\/RI:\ $p\!=\!0.27$; EGFR:\ $p\!=\!0.07$). There is no evidence, therefore, that HPP does not generate statistically identical numbers of reaction firings to both \mbox{NFsim} and SSA. This is as expected since all methods are exact-stochastic approaches.

In Figs.~\ref{fig:tlbr}--\ref{fig:egfr}, panels D, we compare distributions obtained from \mbox{NFsim} and HPP simulations of all models. In Fig.~\ref{fig:tlbr}D, we show equilibrium distributions of the number of receptor clusters in the TLBR model ($f\!=\!0.01$). In Fig.~\ref{fig:actin}D, equilibrium distributions of polymer lengths in the Actin model are shown ($f\!=\!0.01$). In both cases, the \mbox{NFsim} and HPP distributions are statistically indistinguishable (TLBR:\ $p\!=\!0.50$; Actin:\ $p\!=\!0.66$). In Fig.~\ref{fig:fceri}D, time courses for $\gamma$-phosphorylated receptor and receptor-recruited, $\alpha$-phosphorylated Syk are shown ($f\!=\!0.01$). In Fig.~\ref{fig:egfr}D, time courses for membrane-recruited (active) SOS and nuclear phospho-ERK are shown ($f\!=\!0.05$). Although we did not perform any statistical tests, visual inspection of the trajectories clearly shows that in all cases the \mbox{NFsim} and HPP results are virtually identical.

\subsubsection*{Systematic approach to selecting population species}

All of the HPP results presented in Figs.~\ref{fig:tlbr}--\ref{fig:egfr} were obtained with ``hand-picked" sets of population species chosen based on modeler experience and intuition. The significant memory savings seen in these plots imply that this approach will often be sufficient in practice. However, it is fair to ask whether a more systematic approach to selecting population species can achieve additional memory savings. In order to address this question, we considered a variety of different lumping sets for each example model and compared their performance in terms of memory usage and CPU run time. The lumping sets were chosen based on average species populations calculated over the course of a single NFsim pre-simulation at cell fraction $f=0.01$. Specifically, at periodic intervals, the full set of complexes in the system was collected, each complex canonically labeled, and the number of instances of each label (i.e., species) counted. Average values over the entire simulation were then calculated for each species. Sets of population species were constructed by lumping all species with an average population greater than a range of pre-defined thresholds. For convenience, we chose thresholds of $2^n$\/, $n\in[0,10]$. Average species populations obtained from each NFsim pre-simulation are provided in supplementary Dataset~S1. The script that implements this method (for a single threshold) has been included in the recent \mbox{BioNetGen}~2.2.5 release (\texttt{auto\_hpp.pl} in the \texttt{Perl2} subdirectory).

\begin{figure*}[!t]
    \centering
    \caption{
        \textbf{HPP performance analyses for various lumping thresholds at cell fraction {\boldmath$f=0.01$}.}
        (A) TLBR; (B) Actin; (C) Fc$\epsilon$\/RI; (D) EGFR. In all plots, threshold values for different 
        lumping sets are shown on the x-axis.  For TLBR and Actin, some thresholds yielded the same set of population 
        species as larger thresholds and, hence, are omitted from the figures. For TLBR, results for thresholds $<16$\/ are 
        omitted due to impractically large partial networks in those cases. Results for NFsim (``NF") and the hand-picked 
        lumping sets from  Figs.~\ref{fig:tlbr}--\ref{fig:egfr} (``HPP") are shown in all plots for comparison. Error bars show 
        standard error (three samples).
    }
    \label{fig:lumpingSets}
    \includegraphics[width=0.9\textwidth]{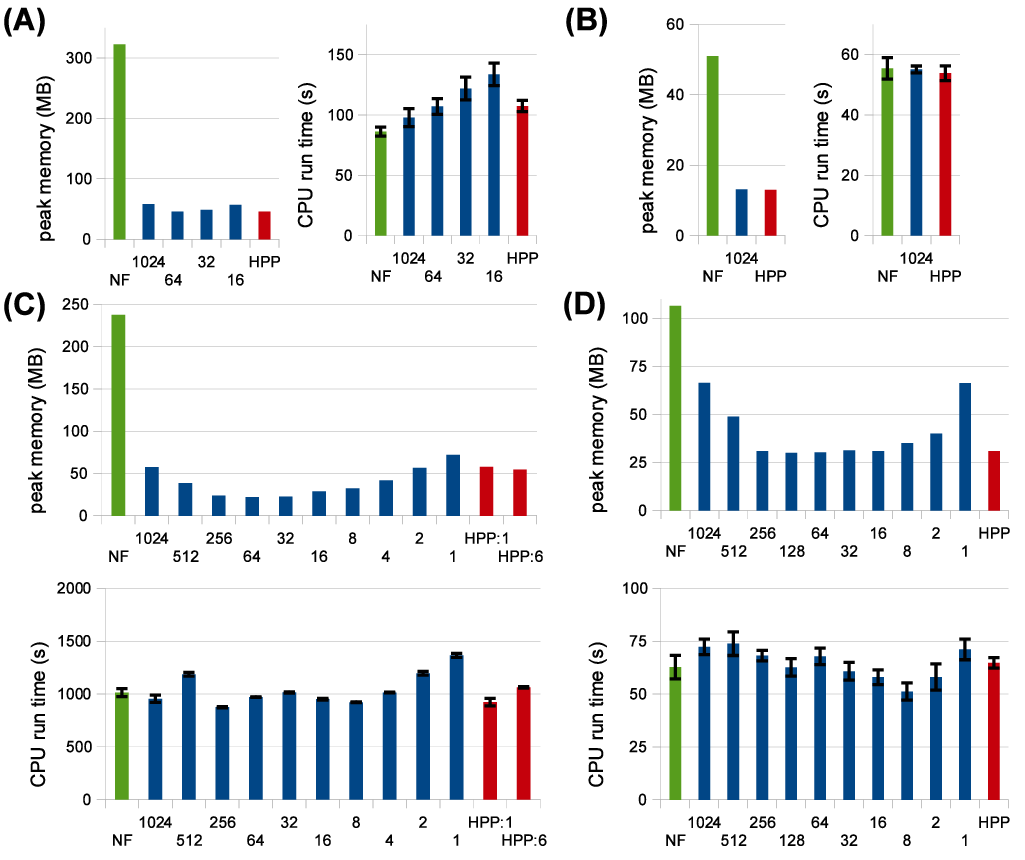}
\end{figure*}

In Fig.~\ref{fig:lumpingSets}, we show peak memory use and CPU run times for HPP simulations of each model at each lumping set considered. In general, these results illustrate the success of the hand-picked lumping sets, which produced memory savings close to the optimal in most cases. There was, however, some room for improvement in the Fc$\epsilon$\/RI model (Fig.~\ref{fig:lumpingSets}C). This is because the fourth and fifth most populated species for this model were complexes comprised of five molecular subunits (see Dataset~S1). Since we did not anticipate this result, these high-population species were not included in the hand-picked lumping set. The majority of the memory savings seen in Fig.~\ref{fig:lumpingSets}C for thresholds $>32$ are due to lumping of these species. Thus, our results also illustrate the value of using a more systematic approach to selecting population species in some cases.

It is also interesting to note in Figs.~\ref{fig:lumpingSets}C and \ref{fig:lumpingSets}D the presence of an optimal lumping threshold between the maximum and minimum values considered. At high thresholds, most species are treated as particles and higher memory use is expected. At low thresholds, however, the higher memory use is due to the larger size of the partially-expanded network. Also interesting is that the run time results in Fig.~\ref{fig:lumpingSets} show a weak (if any) dependence on the chosen threshold, despite the fact that the time complexity of network-free methods scales linearly with rule set size (Table~\ref{table:complexity}). Presumably, this is because the lower cost operations (increment/decrement) associated with the population species offset the increased cost of larger rule sets. This robustness of the time cost with respect to the size of the lumping set is a positive attribute of the HPP method.

\section*{Discussion}

We have presented a hybrid particle/population simulation approach for rule-based models of biological systems. The HPP approach is applied in two stages (Fig.~\ref{fig:hppWorkflow}): (i) transformation of a rule-based model into a dynamically-equivalent hybrid form by partially expanding the network around a selected set of population species; (ii) simulation of the transformed model using a population-adapted network-free simulator. The method is formally exact for an infinite population lumping rate constant, but can produce statistically exact results in practice provided that a sufficiently large value is used (Figs.~\ref{fig:tlbr}--\ref{fig:egfr}, panels C and D). As currently implemented, the primary advantage of the HPP method is in reducing memory usage during simulation (Figs.~\ref{fig:tlbr}--\ref{fig:egfr}, panels A). Importantly, this is accomplished with little to no impact on simulation run time (Figs.~\ref{fig:tlbr}--\ref{fig:egfr}, panels B). 

We have shown that peak memory use for HPP scales linearly with particle number (with a slope that is smaller than for \mbox{NFsim}; Figs.~\ref{fig:tlbr}--\ref{fig:egfr}, panels A) and confirmed that when network generation is possible SSA memory use is approximately independent of particle number (Figs.~\ref{fig:fceri}A and \ref{fig:egfr}A). At the system volumes that we have considered here, HPP memory use is significantly less than for SSA. However, the linear scaling of HPP and the constant scaling of SSA indicate that with further increases in the system volume there will invariably come a point where HPP memory use exceeds that of SSA. This is because species that are rare at small volumes, and hence chosen to be treated as particles, become plentiful at large volumes. Intuitively, a partially-expanded network should never require more memory than a fully-enumerated network. However, as currently implemented, there is no way to strictly enforce this restriction because HPP requires that population species be chosen prior to PNE.

\begin{figure}[!t]
\caption{\textbf{Memory use vs.\ simulated volume for different simulation methods, including a hypothetical automated HPP (aHPP).} For finite networks, aHPP memory use plateaus once the entire reaction network has been generated. For infinite networks, the scaling at large volumes falls somewhere between constant and linear (no worse than HPP) depending on the model (see Sec.~S2 of Text~S1 for an analysis).}
\label{fig:aHPP}
\begin{center}
\includegraphics[width=0.4\textwidth]{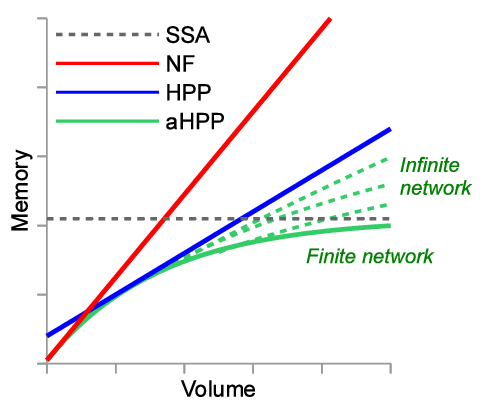}
\end{center}
\end{figure}

In Fig.~\ref{fig:lumpingSets}, we have shown how a systematic approach to choosing population species can optimize memory usage for a given system volume. However, this approach requires running an NFsim pre-simulation, which may not be feasible for systems with extremely large numbers of particles (e.g., whole cells). Thus, we propose to develop a more general version of HPP that dynamically tracks the populations of species during the course of a simulation and automatically selects those to treat as population variables based on some criteria, e.g., that their population exceeds a certain threshold. In this automated version of HPP (aHPP), PNE would be performed every time a new species is lumped. If all species in the system become lumped then the network will naturally become fully enumerated. Hence, the memory load will never exceed that of the fully-expanded network. In Fig.~\ref{fig:aHPP}, we provide a qualitative sketch of how we expect the memory usage of this hypothetical aHPP method to scale with system volume (particle number). Included for comparison are scalings for HPP, \mbox{NFsim}, and SSA. For models with finite networks (such as Fc$\epsilon$\/RI and EGFR), aHPP memory use should plateau once the entire reaction network has been generated. For models with infinite networks (such as TLBR and Actin), we expect aHPP memory use at large volumes to scale somewhere between constant and linear (no worse than HPP) depending on the model. A detailed analysis of the space complexity of a hypothetical, ``optimal" aHPP method is provided in Sec.~S2 of supplementary Text~S1.

\begin{figure}[!t]
\caption{\textbf{Cost of running simulations on the Amazon Elastic Compute Cloud (EC2).} The minimum cost as a function of memory requirement was calculated based on January 2012 pricing (\url{http://aws.amazon.com/ec2/}) of all \emph{Standard\/}, \emph{High-CPU\/}, and \emph{High-Memory\/} EC2 instances (see \textcolor{black}{Sec.~S1 of Text~S1} for details of the calculation). Also included are values for \mbox{NFsim}, HPP, and SSA simulations of the EGFR model at cell fraction $f\!=\!1$.}
\label{fig:memoryCost}
\begin{center}
\includegraphics[width=0.4\textwidth]{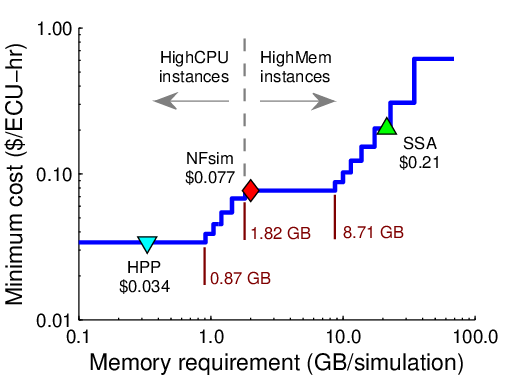}
\end{center}
\end{figure}

In order to frame our results within a real-world context, we have estimated the cost of simulation based on hourly rates of on-demand instances on the Amazon Elastic Compute Cloud (EC2). In Fig.~\ref{fig:memoryCost}, we show the hourly cost (per ``effective compute unit") of simulation as a function of required memory per simulation (details of the calculation can be found in Sec.~S1 of Text~S1). Also included in the plot are values for HPP (0.3~GB), \mbox{NFsim} (2.1~GB), and SSA (22.0~GB) simulations of the EGFR model at cell fraction $f\!=\!1$ (Fig.~\ref{fig:egfr}A). Our calculations show that below 1.82~GB of required memory \textit{High-CPU\/} instances are the most cost effective. Above this threshold \textit{High-Memory\/} instances are the better option. The HPP simulation falls below this cutoff while both \mbox{NFsim} and SSA lie above. There is a quantifiable benefit, therefore, to reducing memory usage in this case; HPP simulations on the EC2 would be $\sim$2.5 and $\sim$33 times less expensive, respectively, than \mbox{NFsim} and SSA (HPP is slightly faster than \mbox{NFsim} and significantly faster than SSA; Fig.~\ref{fig:egfr}B). Thus, the reduction in memory usage offered by HPP is not simply of academic interest but can impact, in a tangible way, the cost of doing computational research.
%

Finally, even greater benefits are possible if, in addition to reducing memory usage, the speed of HPP simulations can be increased. $\tau$\/~leaping \cite{Gillespie2001, Gillespie2003, Cao2006, Gillespie2007} is an approach for accelerating stochastic simulations of chemically reactive systems. With a few exceptions (e.g., Ref.~\cite{Vlachos2008}), $\tau$\/~leaping has been applied primarily to fully-enumerated reaction networks. We believe that the HPP method provides a unique setting for the application of $\tau$\/-leaping because, unlike in pure particle-based methods, there exists a partial network of reactions that act on population species. Thus, a network-based $\tau$\/-leaping method can be applied exclusively to the population component of a system while retaining the network-free approach in the particle component. We have recently implemented a $\tau$\/-leaping variant  in \mbox{BioNetGen}, known as the partitioned-leaping algorithm \cite{Harris2006}, and are actively working on integrating it with the HPP.

\section*{Supporting Information}

\noindent\textbf{Dataset~S1.}\hspace{6pt} Average species populations from NFsim pre-simulations ($f=0.01$) of all example models considered in Fig.~\ref{fig:lumpingSets}.

\noindent\textbf{Figure~S1.}\hspace{6pt} Average number of reactions that must be updated after each reaction firing (i.e, dependencies) for a collection of Fc$\epsilon$\/RI signaling models of varying network size (all models are included in the BioNetGen~2.2.5 release available at \url{http://bionetgen.org}).
%

\noindent\textbf{Text~S1.}\hspace{6pt} \underline{Sec.~S1}: Details of the monetary cost analysis shown in Fig.~\ref{fig:memoryCost}; \underline{Sec.~S2}: Space complexity analyses for the network-based SSA, network-free, HPP, and hypothetical aHPP methods (Fig.~\ref{fig:aHPP}); \underline{Sec.~S3}: Overview of BNGL, model files, and running HPP simulations with \mbox{BioNetGen}/\mbox{NFsim}; \underline{Sec.~S4}: BNGL formalism and the formal foundation of the PNE algorithm (with pseudocode).

\noindent\textbf{Text~S2.}\hspace{6pt} Complete BNGL file for the simple receptor activation model of Fig.~\ref{fig:recAct} (\texttt{receptor\_activation.bngl}).

\noindent\textbf{Text~S3.}\hspace{6pt} HPP configuration file for the simple receptor activation model, including population mapping rules and instructions for executing NFsim and HPP simulations (\texttt{run\_receptor\_activation.bngl}).

\noindent\textbf{Text~S4.}\hspace{6pt} Partially-expanded (HPP) version of the simple receptor activation model of Fig.~\ref{fig:recAct} generated using the method outlined in Fig.~\ref{fig:pneExample} (\texttt{receptor\_activation\_hpp.bngl}).

\noindent\textbf{Text~S5.}\hspace{6pt} BNGL file for the TLBR model (\texttt{tlbr.bngl}).

\noindent\textbf{Text~S6.}\hspace{6pt} HPP configuration file for the TLBR model (\texttt{run\_tlbr.bngl}).

\noindent\textbf{Text~S7.}\hspace{6pt} HPP version of the TLBR model (\texttt{tlbr\_hpp.bngl}).

\noindent\textbf{Text~S8.}\hspace{6pt} BNGL file for the Actin model (\texttt{actin\_simple.bngl}).

\noindent\textbf{Text~S9.}\hspace{6pt} HPP configuration file for the Actin model (\texttt{run\_actin\_simple.bngl}).

\noindent\textbf{Text~S10.}\hspace{6pt} HPP version of the Actin model (\texttt{actin\_simple\_hpp.bngl}).

\noindent\textbf{Text~S11.}\hspace{6pt} BNGL file for the Fc$\epsilon$\/RI model (\texttt{fceri\_gamma2.bngl}).

\noindent\textbf{Text~S12.}\hspace{6pt} HPP configuration file for the Fc$\epsilon$\/RI model (\texttt{run\_fceri\_gamma2.bngl}).

\noindent\textbf{Text~S13.}\hspace{6pt} HPP version of the Fc$\epsilon$\/RI model with free ligand treated as the only population species (\texttt{fceri\_gamma2\_hpp1.bngl}).

\noindent\textbf{Text~S14.}\hspace{6pt} HPP version of the Fc$\epsilon$\/RI model with free ligand, cytosolic Lyn and 
all four phosphorylation states of cytosolic Syk treated as population species (\texttt{fceri\_gamma2\_hpp6.bngl}).

\noindent\textbf{Text~S15.}\hspace{6pt} BNGL file for the EGFR model (\texttt{egfr\_extended.bngl}).

\noindent\textbf{Text~S16.}\hspace{6pt} HPP configuration file for the EGFR model (\texttt{run\_egfr\_extended.bngl}).

\noindent\textbf{Text~S17.}\hspace{6pt} HPP version of the EGFR model (\texttt{egfr\_extended\_hpp.bngl}).

\begin{acknowledgments}
We thank Michael Sneddon for assistance in preparing the \mbox{NFsim}~1.11 distribution (population-adapted).
\end{acknowledgments}


\end{document}